\documentclass[conference]{IEEEtran}

\makeatletter

\usepackage{blindtext}

\usepackage{eso-pic}
\IEEEoverridecommandlockouts

\usepackage{cite}
\usepackage{amsmath,amssymb,amsfonts}
\usepackage{algorithmic}
\usepackage{graphicx}
\usepackage{textcomp}
\usepackage{xcolor}
\def\BibTeX{{\rm B\kern-.05em{\sc i\kern-.025em b}\kern-.08em
    T\kern-.1667em\lower.7ex\hbox{E}\kern-.125emX}}
\definecolor{orcidlogocol}{HTML}{A6CE39}
\usepackage{eso-pic}

\begin{document}
\title{\vspace*{1cm} The Effect of Frequency Droop Damping on System Parameters and Battery Sizing During Load Change Condition\\

}

\author{\IEEEauthorblockN{Mohammed F. Allehyani}
\IEEEauthorblockA{\textit{dept. of Electrical Engineering} \\
\textit{University of Tabuk}\\
Tabuk, KSA \\
mallehyani@ut.edu.sa 
}
\and
\IEEEauthorblockN{Mohamed Abuagreb}
\IEEEauthorblockA{\textit{dept. of Electrical and Computer Eng.} \\
\textit{Clemson University}\\
Clemson, USA \\
mabuagr@clemson.edu}
\and
\IEEEauthorblockN{Brian K. Johnson}
\IEEEauthorblockA{\textit{dept. of Electrical and Computer Eng.} \\
\textit{University of Idaho}\\
Moscow, USA \\
bjohnson@uidaho.edu}
}

\maketitle

\begin{abstract}
Inverter-based resources (IBR) have been widely studied for their advantages on the current power systems. This increase in the penetration of renewable energy has raised some concerns about the stability of the existing grid. Historically, power systems are dominated by synchronous generators that can easily react to system instability due to high inertia and damping characteristics. However, with IBR, the control of the inverter plays a crucial role in contributing to the system stability and enhancing the functionality of the inverters. One of these novel control methods is droop control. Droop characteristics are used to control voltage, frequency, and active and reactive power. This paper presents the impact of frequency droop damping on system frequency, real power, and the rate of change of frequency with distributed energy resources. Also, battery sizing is suggested based on the results. The results also show the need for optimal selection for the frequency droop damping to fulfill the appropriate battery size in terms of cost and performance. The simulations are carried out in an electromagnetic transient program (EMTP).
\end{abstract}


\section{Introduction}
Renewable energy resources (RES) are starting to displace the coal-fired generators in response to government policies to reduce CO2 and other greenhouse gases emissions \cite{9791980}. Wind and photovoltaic (PV) generation are becoming well established, with the deployment of battery energy storage systems (BESS) becoming a common research topic that is moving into industry practice. These technologies are connected to the power grid through power electronics-based voltage source converters (VSC) that also serve to provide decoupling \cite{8397997}. To increase the deployment of RES, the controls for RES need to include functions to mitigate the negative impact of present RES controls and provide an autonomous regulation for grid voltages and frequency. One solution to this problem is using grid-forming inverter controls \cite{8980733}. 

Inverter-based resources can be modeled with grid- forming or following inverter. In grid-forming control, inverters act as voltage sources where the inverter can respond instantaneously to system changes \cite{9683125}. This type of inverter can adjust the voltage magnitude and frequency of the system, supply fault current, and contribute to the system inertia \cite{8980733}. On the other hand, grid-following inverters rely on the measurements of the phase-locked loop (PLL) and the proportional and integral control. Unlike grid-forming inverters, grid-following inverters are limited to tracking the voltage and frequency during disturbances \cite{9683125}.

Different researchers have discussed the concept of droop control in power systems. The authors in \cite{9208657} proposed a droop control based on RoCoF which support the frequency change by controling the DC side capacitor.
A combined droop and virtual inertia control study are proposed in reference \cite{9281722} for a wind farm that operates under underload and overload conditions.  The authors investigate the role of droop coefficient setting in those loading conditions. An intelligent droop control for voltage and frequency regulation is proposed in \cite{6512072}. A comprehensive study about sizing a microgrid considering droop control scheme was conducted in \cite{REY2022107634}. The study reveals that the optimal droop control parameters can be achieved if we design the parameters based on the microgrid sizing, and as a result, a reduction in total cost was successfully obtained.

The focus of this paper is using frequency droop damping control for frequency regulation and inertia support. The effect of different drop damping coefficients on system parameters are investigated. The rest of the paper is structured as follows. In
Section II, an explanation of the droop control in the power system, is presented. Section III describes the system under study and the VSC model developed in this work. Section IV shows the results of a load change condition on the system parameters for different damping coefficients using time-domain simulations. Section V presents discussions and evaluations of the results. Finally, Section VI discusses the outlook of the study and concludes the paper.

\section{Droop Control in Power Systems}
There are several variants of droop control that have been applied in power systems to improve power sharing and
stability such as P\&V droop control, and P\&W droop.

\subsection{Droop Controller}

Droop control is a method used for the control of distributed power that mimics operation of conventional SG behavior. It is used in load sharing and main regulation of the frequency, while the emulated behavior of the virtual inertia is not considered when you model the droop control. The advantages of the droop control is improvement stability of the voltage. Generally, droop control is used when it is assumed that the DC link has large storage. The voltage capacitor at the DC link could oscillate without contribution from storage energy, which causes problems in the stability of the power system. In addition, there is a concern about the voltage at the DC capacitor to reach the reference value, when tracking the MPPT of the PV generation. Thus, another additional control loop of the voltage should be combined with the droop controller loop to regulate the power output from PV inverter and then regulate the voltage at the DC capacitor. 

There are two implemented controllers; either can be used in the droop controller. The first controller is essentially a combination of the traditional control of power converter and an additional block used to adjust reference of the active power for the converter through relative action that uses its input for the deviation of the frequency. In this approach, the converter that feeds the grid with the controllable for both, reactive and active powers, can be a grid-aiding converter. The power reference that is obtained from the outer loop is shown in figure \ref{P3,7}.\cite{Ref3.2}, \cite{Ref3.3} and\cite{Ref2.7}

\begin{figure}[htbp!]
	\centering
\centerline{\includegraphics[scale=0.6]{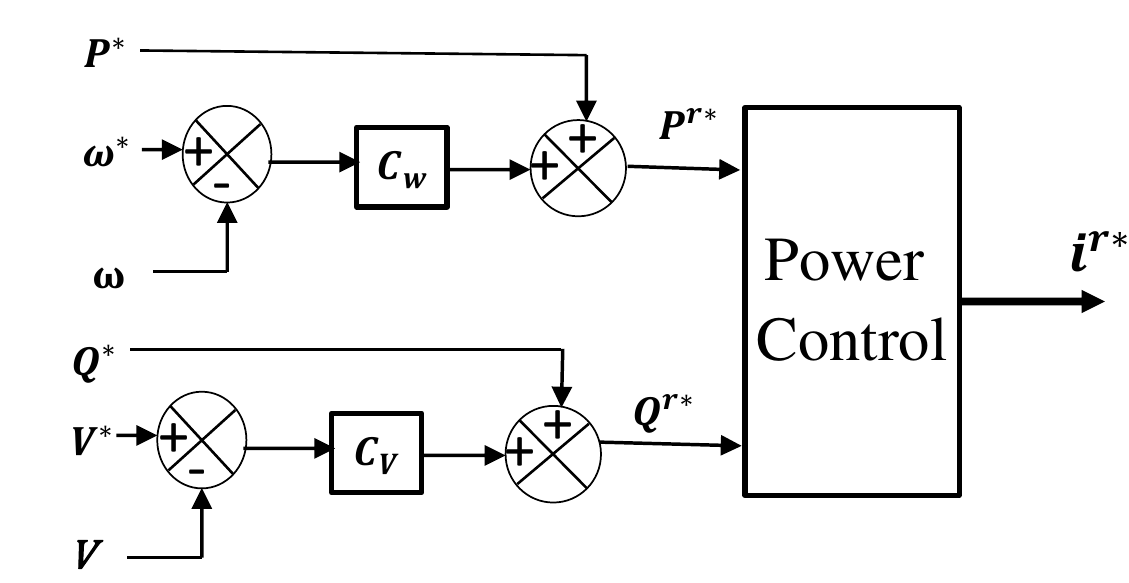}}
	\caption{Traditional droop controller depends on converter of the grid-feeding  } 
	\label{P3,7}
\end{figure}

The other approach shown in figure \ref{P3,8} shows that the converter can be seen as a voltage source with a frequency, which is calculated from the power error. As a result, the internal loop of the traditional converter will reproduce the voltage after the virtual impedance has improved it. In addition, the frequency and the angle of the voltage source are used for synchronization. By using this method, there is no need to add any system's synchronization, and the model will be defined as an impedance behind the voltage source \cite{Ref2.7}.

\begin{figure}[htbp!]
	\centering
\centerline{\includegraphics[scale=0.6]{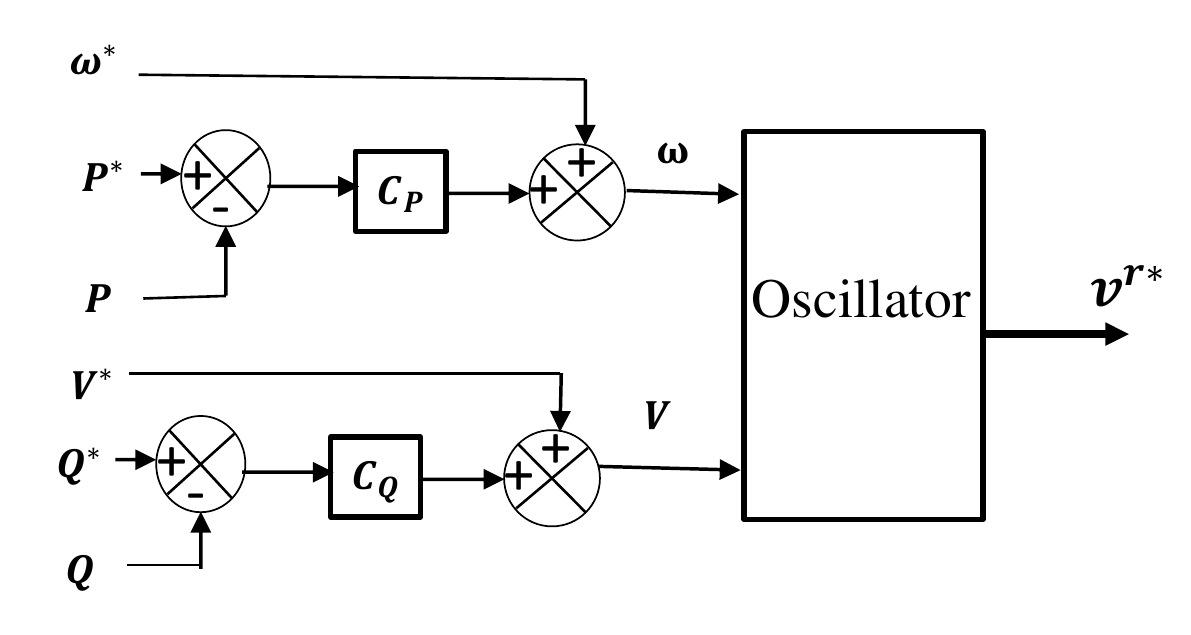}}
	\caption{Traditional droop controller depends on converter of the grid-feeding  } 
	\label{P3,8}
\end{figure}

\subsection{Damping Controller}

The result of the analysis conducted on the synchronous machines' dynamics which is based on the swing equation without damping reveals the marginal stability of the response of the active power.  This is why synchronous machines are often designed with damper windings, although this impacts the effectiveness of the machine as the losses increase. Different damping values were however observed due to the peculiarity of the controllers studied. Take for example the case of the synchronverter, the damping coefficient is computed as a function of the preferred steady state droop due to the absence of an external loop. This often causes the damping ratio for the closed loop to be very low. The actual grid voltage is usually used to determine the damping of other virtual synchronous machines (like VSG with direct modulation). In this case, frequency regulation is achieved by a separate external droop. The advantage of this setup is that it gives room for easy decoupling of the steady-state and the transient responses; hence, the damping coefficient defined can be as large as possible without affecting the machine's steady-state behavior. Effectiveness may however be impacted negatively when oscillation is involved \cite{Ref3.2} and \cite{Ref2.7}.

\section{System Configuration and Control}
\begin{figure}[htbp!]
\centerline{\includegraphics[scale=0.5]{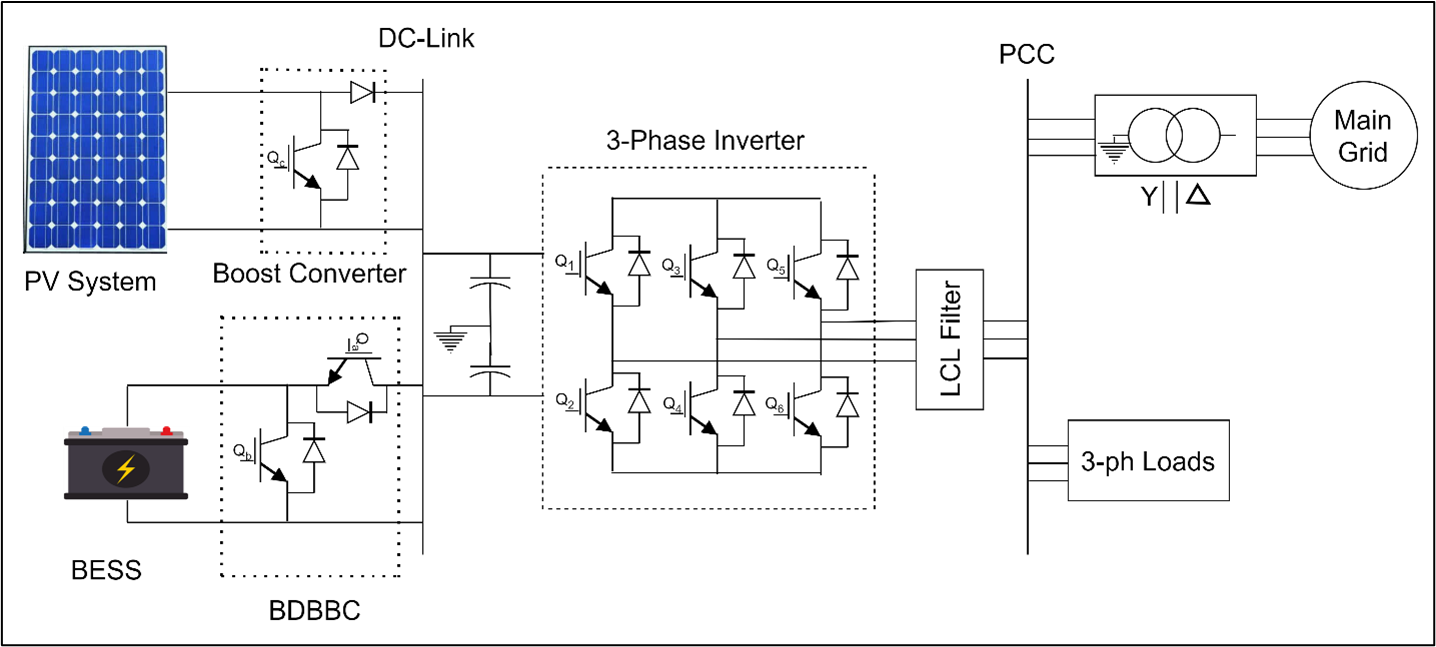}}
\caption{Simplified diagram of the overall model.}
\label{model}
\end{figure}

 The system in Fig. \ref{model} is built in the EMTP-ATP software. The model is divided into three main parts which are: The DC system, the AC system, and the voltage source converter (VSC). First, the DC system consists of the PV and BESS systems. The PV system is fully modeled based on specific irradiation and temperature conditions. Also, a MPPT control mechanism is designed based on the well-known P\&O algorithm. The maximum
power from the PV is fed to the load via a boost converter to step up the voltage to the required magnitude. The BESS is modeled as a lithium-ion battery and works as a controlled voltage source.
The Battery is connected to the DC-link through a bidirectional buck-boost converter (BDBBC). The AC system consists of three phase loads and the step up transformer that is connected to the AC system. A two level three phase VSC is modeled as an average model with grid-forming converter functions.

\subsection{PV Model}\label{AA}
PV arrays consist of parallel and series combinations of cells to produce the required voltage and current. The voltage and current behavior of the PV array can be expressed using nonlinear mathematical equations described in \cite{7771689}

A table description of the MPPT can be found in table \ref{PV}

\renewcommand{\arraystretch}{1.2}
\begin{table}[htbp!]
	\centering
		\caption{The parameters of the KC200GT solar array at $25^oC$, $1000 W/m2$ [8]}
		\vspace{-0.1 cm}
\begin{tabular}{|l|l|l|l|}
	\hline Parameters & Symbol & Value \\
	\hline
	Voltage-temperature coefficient & $K_V$& -0.123v/K\\
	\hline
	Current-temperature coefficient & $K_i$& 0.0032A/K \\
	\hline
	Open-circuit voltage at $T_n$ and $G_n$ & $V_{0cn}$ & 32.9V\\
	\hline
	Short-circuit current at $T_n$ and $G_n$ & $I_{scn}$ & 8.2 A\\
	\hline
	Experimental maximum power& $P_{max,e}$ & 200.14W\\
	\hline
	Current at maximum power & $I_{mp}$ & 7.6 A\\
	\hline
	Voltage at maximum power & $V_{mp}$ & 26.3\\
	\hline
	Parallel numbers of PV modules & $N_p$ & 500\\
	\hline
	Serious numbers of PV modules & $N_s$ & 24\\
	\hline

	\end{tabular}\\
\label{PV}
\end{table}

\subsection{Battery Energy Storage Model}

The equivalent circuit model for dynamic simulation is used rather than electrochemical or experimental models. In this work, the generic battery model of Lithium-Ion is used.
the parameters of the is presented in table \ref{tbl:excel-table}.
The inherent equation which applies to the charging and discharging operations of the Lithium-Ion model is given in eq (\ref{eq2}).

 \begin{table}[htbp!]
	\centering
	\caption{The Parameters of the BESS}
	\label{tbl:excel-table}
	\includegraphics[width=8cm,height=3.4cm]{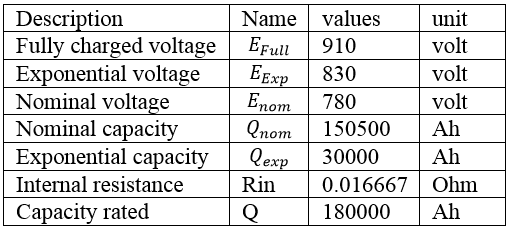}
	\label{table:table51}
\end{table}

\begin{equation}
V_{Batt}=E_{0}-Ri-K(\frac{Q}{(Q-i_t)})i^*+ Aexp(-B.i_t)
\label{eq2}
\end{equation}

 Where $E$ is voltage controller source which is nonlinear $(V)$; $E_0$ is described by no load constant voltage $(V)$; $B$ is exponential capacity  $(Ah )^-1$   ; K  is polarization voltage constant $(V)$ ; $i^*$  is low frequency current dynamics $(A)$ ; $i$  is current of the battery $(A)$ ; $it$  is extracted capacity $(Ah)$;  $Q$ is the capacity of the Battery $(Ah)$; $A$ is exponential of voltage zone amplitude $(V)$. $R$; internal resistance $\Omega$. $V_{Batt}$ is terminal battery voltage.

\subsection{VSC and Frequency Droop Control Models}
\begin{figure}[htbp!]
\centerline{\includegraphics[scale=0.54]{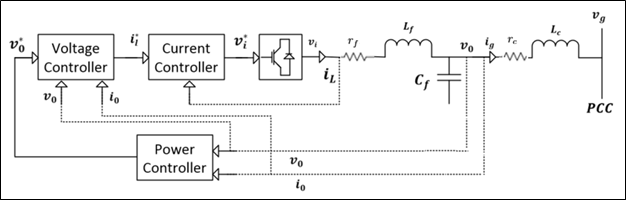}}
\caption{The VSC control design.}
\label{VSC}
\end{figure}

The inverter is built based on a state space average model of a two level three phase voltage source converter (VSC).  The control schemes of the VSC consist of two parts, outer (voltage) and inner (current) controls as shown in Fig. \ref{VSC}.
The first part is an output loop control, which is either real power control or a combination of DC voltage control or an AC voltage control. The second controller is an inner current regulator designed to reject the disturbances from the higher frequency switching and protect the VSC switching device from overcurrent \cite{7010055}.

The droop control is built based on eq \eqref{eq3}.

\begin{equation}\label{eq3}
\dot{\omega}={(P^{*}-P_{out}+K_d(\omega_{r}-\omega_{o})}
\end{equation}


Where $P_{ref}$ is the reference power of the inverter;
$P_{out}$ is the output power of inverter;
$K_d$ damping constant; $\omega_{r}$ is the reference frequency; $\omega_{o}$ is the grid frequency.

\begin{figure}[htbp!]
\centerline{\includegraphics[scale=0.55]{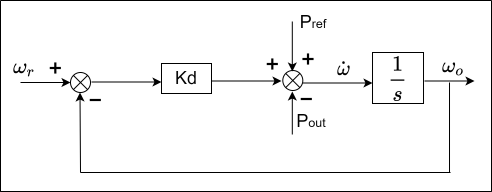}}
\caption{The block diagram of the artificial frequency droop control.}
\label{droop}
\end{figure}

The frequency droop control circuit is applied to the inverter control schemes to provide damping properties during a disturbance. The effect of different damping constants is dis-
cussed in this section. In Fig. \ref{droop}, the damping coefficient (Kd) is varied between almost zero to 140 in order to present the effect of Kd on the rate of change of frequency response and battery size needed in each case.

\section{Simulation Results}
\begin{figure}[htbp!]
\centerline{\includegraphics[scale=0.19]{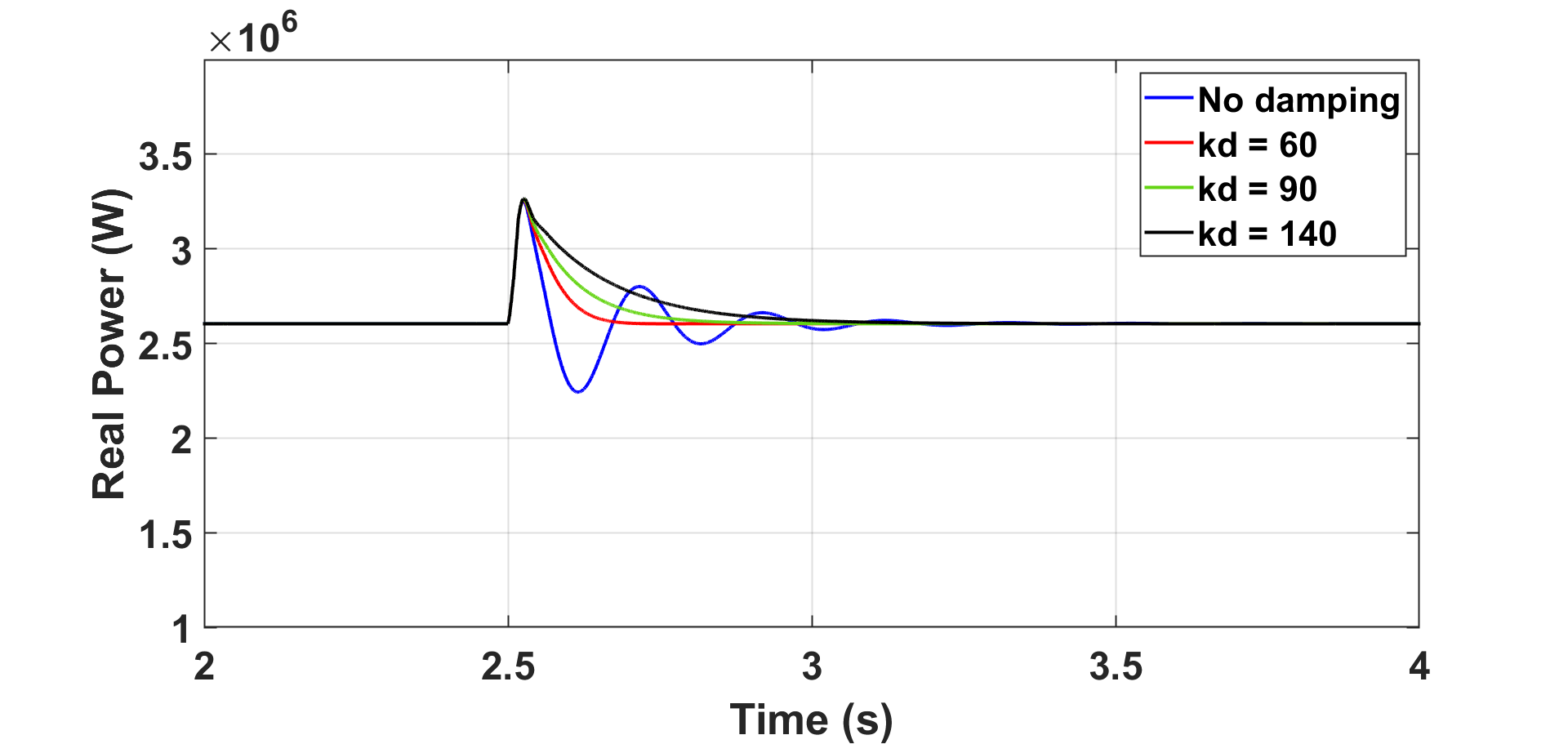}}
\caption{The system's real power during load change.}
\label{p3ph}
\end{figure}

The system in Fig. \ref{model} is tested under load change condition. The simulation is carried out in an electromagnetic transient program (EMTP). The load in Fig.\ref{model} has increased from 2.6 MW to 3.4 MW (the load has increased by 30$\%$) as shown in Fig. \ref{p3ph}. While the load is changed the system parameters,(frequency, RoCoF, the supported power of the battery and the DC link voltage are monitored) 

\begin{figure}[htbp!]
\centerline{\includegraphics[scale=0.19]{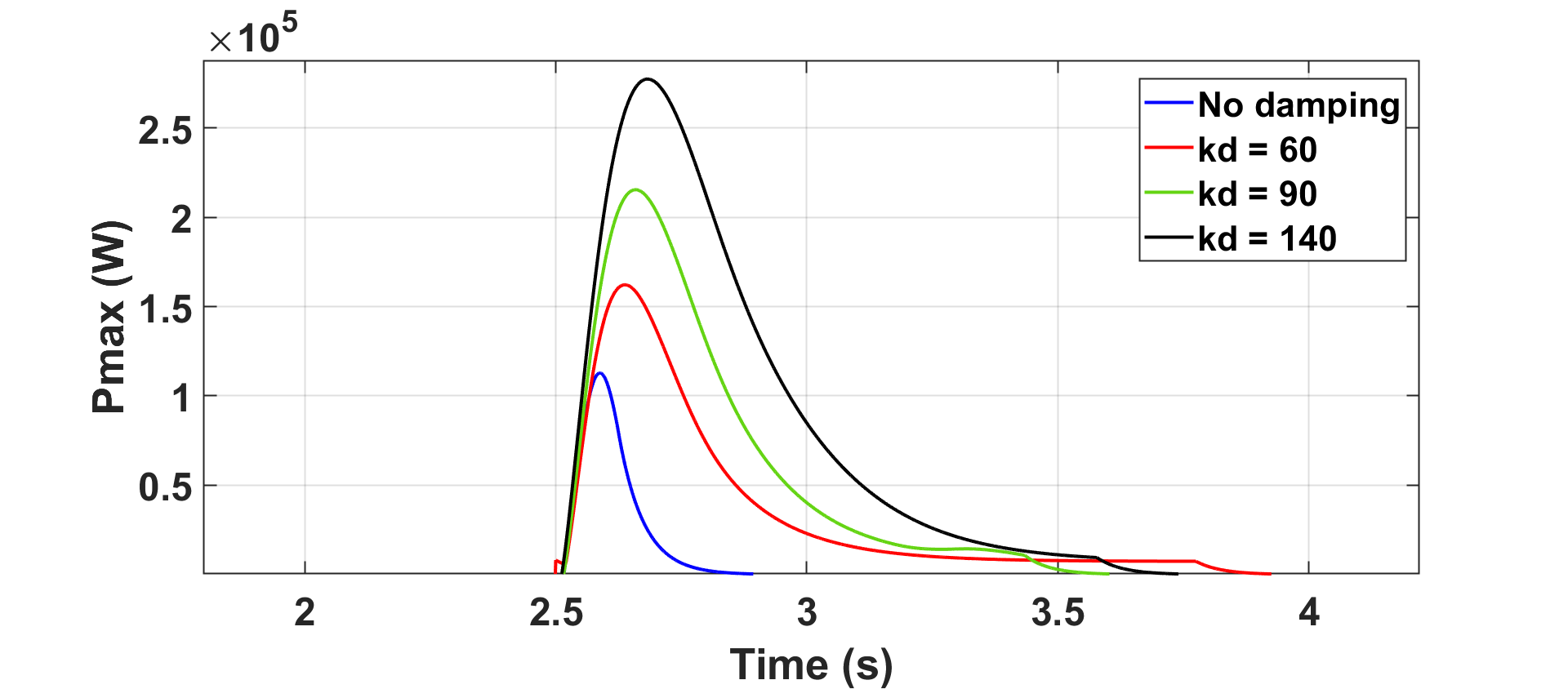}}
\caption{Maximum battery power support for different damping coefficients.}
\label{pbattery}
\end{figure}

\begin{figure}[ht!]
\centerline{\includegraphics[scale=0.19]{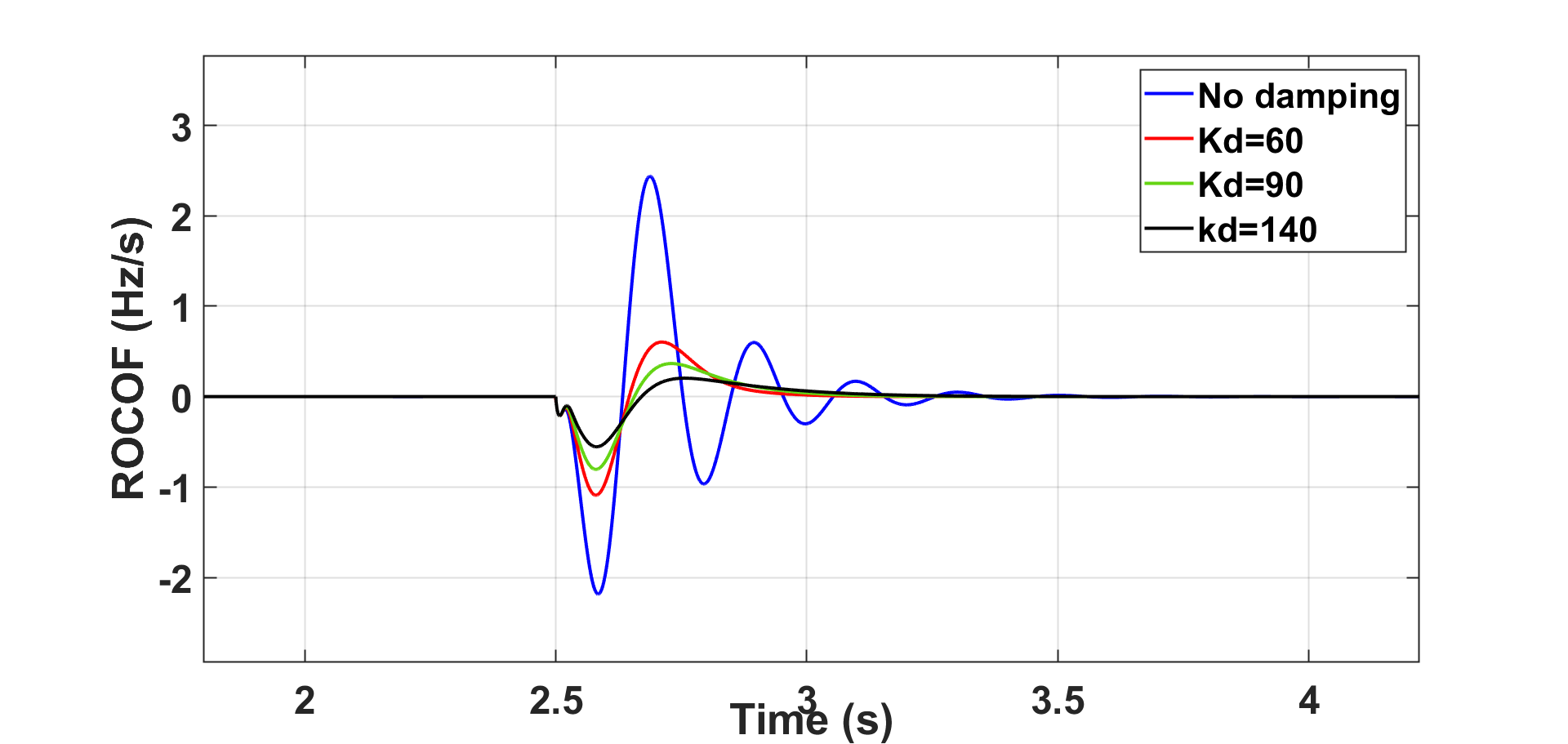}}
\caption{The rate of change of frequency for different damping coefficients.}
\label{rocof}
\end{figure}

The results of Fig. \ref{pbattery} and Fig. \ref{rocof} indicate that the bigger the value of Kd, the smaller the ROCOF. This will help design the appropriate damping constant that does not trigger the frequency protection schemes. However, Fig. \ref{pbattery} shows that the bigger the value of Kd, the bigger the size of the battery needed. Therefore, the trade-off is between performance
and cost, which is a critical design decision.

\begin{figure}[ht!]
\centerline{\includegraphics[scale=0.19]{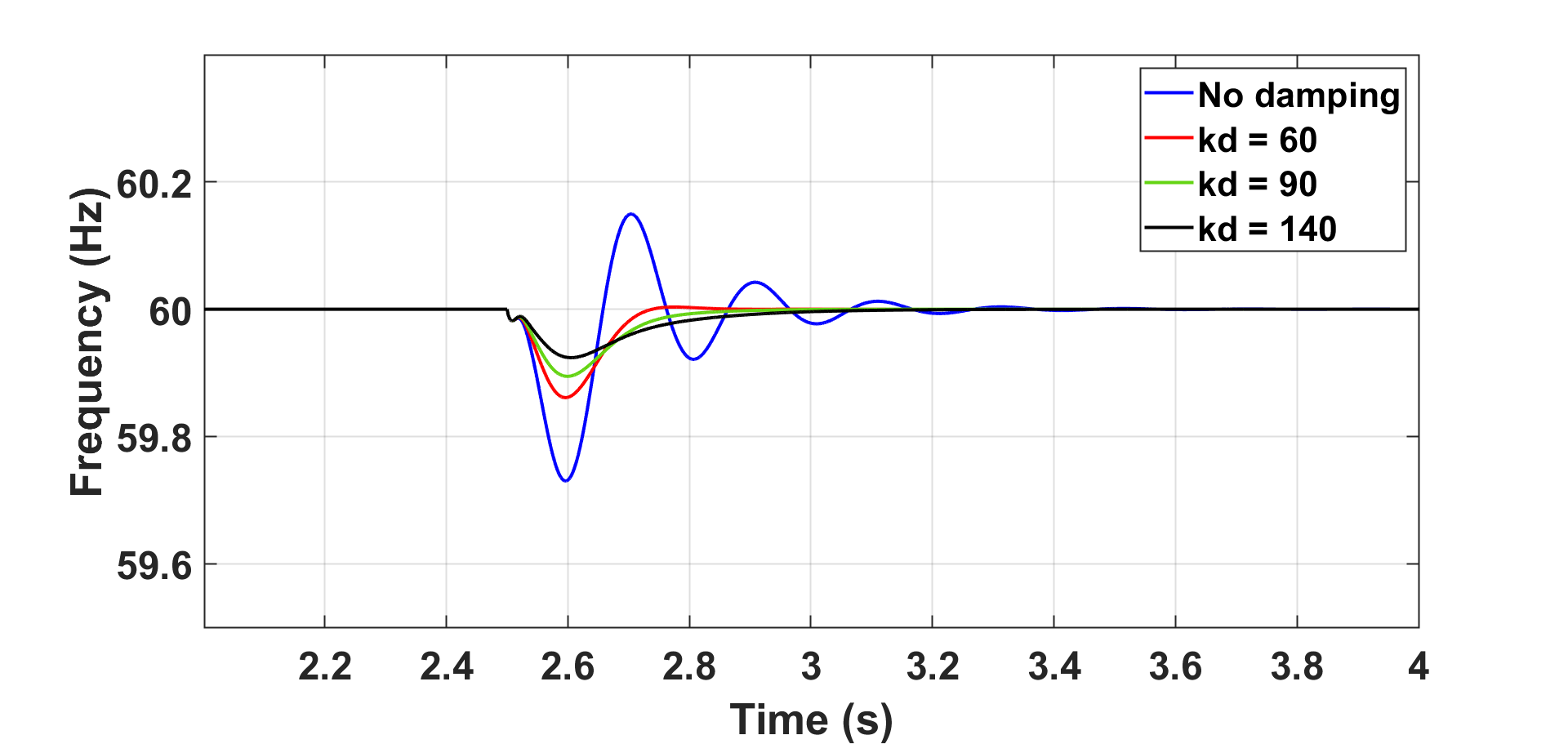}}
\caption{System's frequency response.}
\label{freq}
\end{figure}

The frequency droop control is used as an inertia support control. The results show that droop control provide damping that restored the system in case of disturbance or load change as shown in Fig. \ref{freq}

Under frequency load shedding (UFLS) scheme has to react if the frequency continues to deviate more that violates the UFLS settings and hit the threshold. Therefore, frequency droop and emulated inertia control are suggested to avoid such a condition.

The results show the bigger the damping coefficient, the more significant the overshoot, and the slower the settling time.

\begin{figure}[ht!]
\centerline{\includegraphics[scale=0.19]{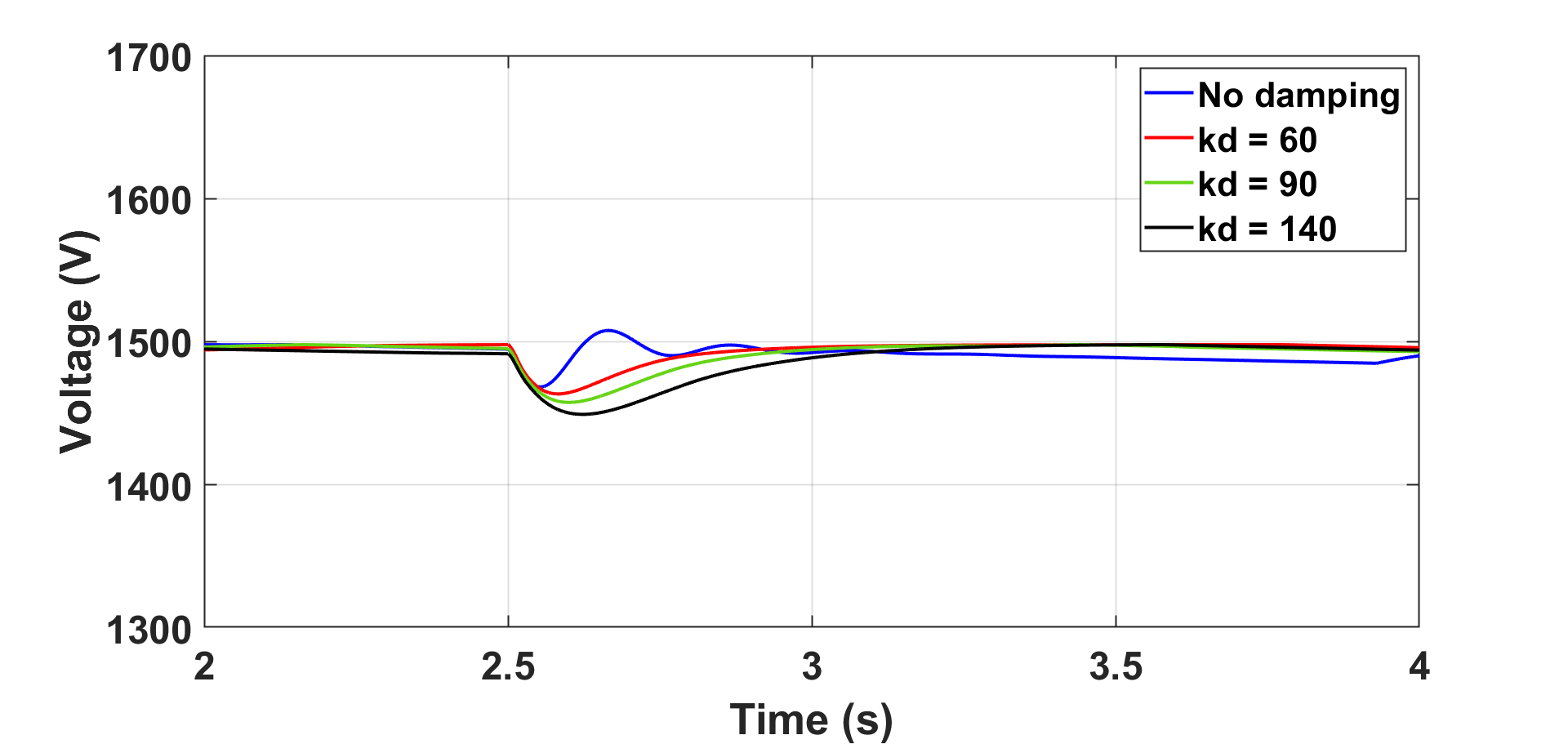}}
\caption{DC link voltage.}
\label{vdc}
\end{figure}

The DC link voltage is set at 1500 V during normal operation. During load change condition, the voltage has slightly affected with different droop damping coefficients as shown in Fig. \ref{vdc}. 

\renewcommand{\arraystretch}{1.0}
  \begin{table}[htbp!]
\caption{Comparison of System Parameters for a Load change.}
\vspace{-0.1 cm}
\begin{tabular}{c c c c c}
\hline
\\Parameters & No Damping &   Kd = 60 & Kd = 90 & Kd = 140\\
\\\hline

\\Freq & 60.15 Hz  & 60 Hz & 59.98.5 Hz& 59.98 Hz\\
\\

\\RoCoF & 2.43  & 0.604& 0.362& 0.204\\

\\

\\Pbattery & 112.8 kW  & 162 kW& 215.4 kW& 277.1 kW\\

\\

\\Vdc & 1508 V & 1464 V& 1458 V& 1449 V\\

\\\hline
\end{tabular}
 \label{tab:summary}
\end{table}

According to the summary of the results in table \ref{tab:summary}, the optimal damping coefficient is suggested to be between (kd= 60 - kd = 90). This is because in this kd range, the frequency is in the normal operating condition, and the RoCoF will not reach the threshold for UFLS.
Higher RoCoF leads to pole slipping which normally occurs when the RoCoF ranges from 1.5 to 2 Hz/s \cite{8506338}.
Also, the DC link voltage will be in an acceptable range, and the battery can be sized so that the battery and the converter will not exceed their capacities.

Depending on Kd, we will have different contributions. The converter will reach the maximum current that the converter can deliver and will reach its maximum power that can be delivered as well.


\section{Conclusion}
In power systems, frequency droop control is used to provide frequency regulation and support system inertia, especially in the presence of inverter-based resources where low inertia is the common case. The motivated advantage of using droop control is the low cost and easy implementation in the power system; however, the droop coefficient plays a vital role in the feasibility of this control technique. In this paper, the role of the damping coefficient of the frequency droop is investigated in time-domain simulations. The system is tested during load change conditions while having different damping coefficients. The results show that choosing a higher or smaller damping coefficient is not necessarily the best option. Therefore, it is important to study a particular system in order to optimally select the best damping constant that enhances the system response while maintaining the system limits in terms of battery and converter sizing. For future studies, optimization studies for selecting the droop damping coefficient need to be investigated in more detail. Also, conducting protection studies is another enhancement for this study to show the response of the under-frequency load shedding scheme for different droop coefficients. Since the droop control method has some drawbacks, which limit its applicability for a modern power system, it is suggested that combining another control technique with the droop control will enhance the overall system behavior.

\bibliographystyle{IEEEtran.bst}
\bibliography{Ref}
\end{document}